\documentclass[twocolumn,showpacs,preprintnumbers,nofootinbib,prd,superscriptaddress]{revtex4-1}
\usepackage{graphicx,amssymb,amsmath,amsthm,amsfonts,epsfig}

\usepackage[linktocpage]{hyperref}
\usepackage[usenames]{color}
\usepackage{epstopdf}

\def\nn{\nonumber}

\newcommand{\ben}{\begin{enumerate}}
\newcommand{\een}{\end{enumerate}}

\def\be{\begin{equation}}
\def\ee{\end{equation}}
\def\bea{\begin{eqnarray}}
\def\eea{\end{eqnarray}}
\newcommand{\beq}{\begin{eqnarray}}
\newcommand{\eeq}{\end{eqnarray}} 

\begin{document}
\title{Black holes with surrounding matter in scalar-tensor theories}

\author{Vitor Cardoso} 
\affiliation{CENTRA, Departamento de F\'{\i}sica, Instituto Superior T\'ecnico, Universidade T\'ecnica de Lisboa - UTL,
Av.~Rovisco Pais 1, 1049 Lisboa, Portugal.}
\affiliation{Perimeter Institute for Theoretical Physics
Waterloo, Ontario N2J 2W9, Canada.}
\affiliation{Department of Physics and Astronomy, The University of Mississippi, University, MS 38677, USA.}

\author{Isabella P. Carucci}
\affiliation{CENTRA, Departamento de F\'{\i}sica, Instituto Superior T\'ecnico, Universidade T\'ecnica de Lisboa - UTL,
Av.~Rovisco Pais 1, 1049 Lisboa, Portugal.}
\affiliation{Dark Cosmology Centre, Niels Bohr Institute, University of Copenhagen, Juliane Maries Vej 30, 2100 Copenhagen, Denmark}

\author{Paolo Pani}
\affiliation{CENTRA, Departamento de F\'{\i}sica, Instituto Superior T\'ecnico, Universidade T\'ecnica de Lisboa - UTL,
Av.~Rovisco Pais 1, 1049 Lisboa, Portugal.}
\affiliation{Institute for Theory $\&$ Computation, Harvard-Smithsonian
CfA, 60 Garden Street, Cambridge, MA, USA.}

\author{Thomas P. Sotiriou}
\affiliation{SISSA, Via Bonomea 265, 34136, Trieste, Italy and INFN, Sezione di Trieste, Italy.}


\begin{abstract} 

We uncover two mechanisms that can render Kerr black holes unstable in scalar-tensor gravity, both associated to the presence of matter in the vicinity of the black hole and the fact that this introduces an effective mass for the scalar. 
Our results highlight the importance of understanding the structure of spacetime in realistic, astrophysical black holes in scalar-tensor theories.
\end{abstract}

\maketitle

The most studied alternatives to general relativity (GR) are scalar-tensor theories (S-T), with action~\cite{Fujii:2003pa,valeriobook}
\be
S=\int d^4x \frac{\sqrt{-g}}{16\pi G}\left(F(\phi)R-Z(\phi)\left(\partial\phi\right)^2+ V(\phi)\right)+S_m,\label{actionST}
\ee
where $R$ is the Ricci scalar of the spacetime metric $g_{\mu\nu}$, $\phi$ is a scalar field, and $S_m$ denotes the matter action. The matter fields $\Psi_m$ are minimally coupled to $g_{\mu\nu}$ and do not couple to $\phi$.
The functionals $F$ and $Z$ single out the theory within the class, up to a degeneracy due to the freedom to redefine the scalar (see, {\em e.g.}~\cite{Sotiriou:2007zu}). 
S-T theory is expected to encapsulate some of the infrared phenomenology of quantum gravity candidates with extra scalar degrees of freedom, such as the dilaton in string theory. For instance, the low-energy limit of bosonic string theory corresponds to $F=\phi$, $Z=-\phi^{-1}$. Brans--Dicke theory \cite{bd} corresponds to $F=\phi$, $Z=\omega_0/\phi$. S-T theories can also be thought of as effective descriptions of a spacetime-dependent gravitational coupling.
They have received widespread interest in cosmology, acting as a rather general parametrization for dark energy~\cite{Clifton:2011jh}.

Of particular interest is the phenomenology in the strong-gravity regime. The reason is twofold: First, it can provide insight on how extra fundamental fields affect the structure of compact stars and black holes (BHs). Second, the study of these objects and the confrontation with observations can yield important constraints on the theory itself~\cite{Will:2005va}. S-T theories  seem to have screening mechanisms that allow the scalar to go undetected in the solar system \cite{Khoury:2003aq,Hinterbichler:2010es}, so strong-gravity constraints can be the ideal way to distinguish them from GR.

In Ref.~\cite{Sotiriou:2011dz} it was shown that asymptotically flat BHs in S-T theory that are stationary (as endpoints of collapse) are no different than BHs in GR in electrovacuum. That is, the scalar field settles to a constant and spacetime is described by the Kerr--Newman family of solutions. This is not to say that BHs cannot be used as probes in order to distinguish S-T theory from GR: the spacetime might be the same, but perturbations will behave differently as the two theories have different dynamics \cite{Barausse:2008xv}. In fact, the existence of a scalar mode in the spectrum of perturbations around a Kerr BH has been shown to lead to remarkable effects \cite{Cardoso:2011xi,Yunes:2011aa}.

Compact stars in S-T have also been studied and an unexpected phenomenon has been discovered: up to a certain density, stars tend to prefer a ``hairless'' configuration. However, above a threshold density ``spontaneous scalarization" occurs and the scalar develops a nontrivial profile~\cite{Damour:1993hw,Damour:1996ke,Pani:2010vc,Barausse:2012da}. Here, we uncover a similar mechanism for BHs with surrounding matter: when the matter configuration is dense enough, the scalar acquires a negative effective mass squared and the BH is forced to develop scalar hair. GR black holes are still solutions of the field equations but are not entropically favoured.

On the other hand, when the effective mass squared of the scalar is positive and the BH spin is sufficiently large, a different kind of instability can occur, due to superradiance~\cite{Press:1972zz}. This instability does {\it not} lead to a non-GR solution, but rather extracts rotational energy away from the BH, which is forced to spin-down.

\noindent{\bf{\em Framework.}}
Action (\ref{actionST}) is said to be written in the Jordan frame. Via the conformal transformation $g^E_{\mu\nu}=F(\phi)g_{\mu\nu}$ and the field redefinition $4\sqrt{\pi} F(\phi) d\Phi=\sqrt{3F'(\phi)^2+2Z(\phi) F(\phi)}d\phi$, one moves to the Einstein frame 
where $\Phi$ is minimally coupled to gravity, but any matter field $\Psi_m$ is coupled to the metric $A(\Phi)^2g_{\mu\nu}^E$
with $A(\Phi)=F^{-1/2}(\phi)$. For what follows we neglect the potential $V(\phi)$, as it is not crucial in our discussion.
The field equations in the Einstein frame read  (setting hereafter $\hbar=c=G=1$)
\begin{eqnarray}
 \label{einein} G_{\mu\nu}^E&=&8\pi \left(T_{\mu\nu}^E+\partial_\mu\Phi\partial_\nu\Phi-{g_{\mu\nu}^E}(\partial\Phi)^2/2\right)\,,\\
 \square^E\Phi&=&-T^E{d [\ln A(\Phi)]}/{d\Phi}\,,
\end{eqnarray}
where $ {T^\mu_\nu}^E=A^4(\Phi) T^{\mu}_{\nu}$.
Expanding around a solution $\Phi_{0}$ to first order in $\varphi\equiv\Phi-\Phi_{0}\ll1$ we obtain~\cite{Yunes:2011aa}
\beq
&&G_{\mu\nu}^E/(8\pi)=T_{\mu\nu}^E+\partial_\mu\Phi_0\partial_\nu\Phi_0-g_{\mu\nu}^E(\partial\Phi_0)^2/2\nonumber\\
&&+\partial_\mu\Phi_0\partial_\nu\varphi+\partial_\mu\varphi\partial_\nu\Phi_0-g_{\mu\nu}^E\partial_\mu\Phi_0\partial^{\mu}\varphi\,, \label{Einsteinlin} \\
&&\square^E\Phi_0+\square^E\varphi=-\alpha_1T^E+\left(\alpha_1^2-2\alpha_2\right)\varphi T^E\,, \label{KGlin}
\eeq
Here, we assumed a general analytical behavior around $\Phi\sim \Phi_{0}$, $A(\Phi)/A(\Phi_0)=\sum_{n=0}\alpha_n(\Phi-\Phi_{0})^n$.

As is obvious from 
Eq.~(\ref{KGlin}), $\alpha_1$ controls the effective 
coupling between the scalar and matter. Various observations, such as 
weak-gravity constraints and tests of violations of the strong 
equivalence principle, seem to require $\alpha_1$ to be negligibly small 
when the scalar takes its asymptotic value \cite{Damour:1998jk,Damour:1996ke,Freire:2012mg}. This implies that a 
configuration in which the scalar is constant and $\alpha_1\approx 0$ is 
most likely to be at least an approximate solution in most viable 
S-T theories. From here onwards we therefore set $\alpha_1=0$, with the understanding that 
in this spirit our analysis and results appear to be rather generic when one restricts 
attention to viable S-T theories.

With $\alpha_1=0$ and a background GR solution all that remains, to first order in $\varphi$, is the Klein-Gordon equation
\be
\left[\square^E-\mu_s^2(x^\nu)\right]\varphi=0\,,\qquad \mu_s^2(x^\nu)\equiv -2\alpha_2 T^E\,. \label{effectivemass}
\ee
Thus, couplings of scalar fields to matter are equivalent to an effective spacetime-dependent mass. 
Depending on the sign of $\alpha_2$ and $T^E$, the effective mass squared can be either positive or negative. Depending on the sign, two types of instabilities, which we detail below, may drive the background solution to develop scalar hairs.

\noindent{\bf{\em Spontaneous scalarization.}}
The most important result of our analysis is that a matter distribution $T^E$ around BHs forces the scalar field to be spontaneously excited and develop a non-trivial configuration. In other words, even though GR is a solution of the field equations, it may not be the entropically
preferred configuration. 
This phenomenon is the direct analog of spontaneous scalarization 
first discussed for compact stars by Damour and Esposito-Far\`ese
\cite{Damour:1993hw,Damour:1996ke,Pani:2010vc,Barausse:2012da}. At 
linear level, spontaneous scalarization manifests itself as a 
tachyonic instability triggered by a \emph{negative} effective mass 
squared. 

Let us first consider the case in which $T^E$ is spherically symmetric, $T^E=T^E(r)$, and its backreaction on the geometry is negligible. In this probe limit the background metric is a Schwarzschild BH. After a decomposition in spherical harmonics $\varphi(t,r,\theta,\phi)=\sum_{lm}e^{-i\omega t}Y_{lm}(\theta,\phi)\Psi_{lm}(r)/r$, the scalar field then obeys 
\beq
&&f^2\Psi_{lm}''+f'f\Psi_{lm}'+\left[\omega^2-f{\cal V}(r)\right]\Psi_{lm}=0\,,\label{KGsphe}\\
&&{\cal V}(r)=\frac{l(l+1)}{r^2}+\frac{2M}{r^3}+\mu_s^2(r)\,,
\eeq
where $f=1-2M/r$ and $'\equiv d/dr$. 
This is
an eigenvalue equation for $\omega=\omega_R+i\omega_I$, when the 
eigenfunctions $\Psi_{lm}(r)$ are required to satisfy appropriate boundary conditions, viz. out-going waves
at spatial infinity, $\Psi_{lm} \sim e^{+i\omega r_*}$ and in-going at the horizon, $\Psi_{lm} \sim e^{-i\omega r_*}$ \cite{Berti:2009kk}. 
Because $\varphi\sim e^{-i\omega t}$, unstable modes correspond to $\omega_I>0$, and they decay exponentially at the boundaries.
In this case, one can make contact with and borrow some powerful results from quantum mechanics. In particular,
a {\it sufficient} condition for this potential to lead to an instability is that
$\int_{2M}^{\infty} {\cal V}dr<0$,
which yields the instability criterion~\cite{Buell}
\be
2\alpha_2\int_{2M}^{\infty} T^E dr>\frac{2l(l+1)+1}{4M}\,.\label{sufficient}
\ee
The above is a very generic, analytic result. We have checked numerically for specific models that the inequality is nearly saturated for interesting matter configurations. For instance, consider $\mu_s^2=-\Theta(r-r_0)\beta M^{n-3}(r-r_0)/r^n$
with $\Theta$ the Heaviside function. This matter distribution, chosen quite arbitrarily to make our point, models the existence of an innermost-stable circular orbit close to the event horizon by not allowing matter to be closer than $r=r_0$.
Spontaneous scalarization occurs for
\be
-\alpha_2\frac{\mu}{M} \gtrsim \pi \left(2l(l+1)+1\right)\frac{(n-2)(n-1)}{(n-4)(n-3)}\left(\frac{r_0}{M}\right)^2\,,\label{model1}
\ee
where $\mu=-4\pi\int r^2 T^E$ is the mass of the spherical distribution and its finiteness requires $n>4$.
A minimum mass $\mu$ is thus necessary in order for 
spontaneous scalarization to occur. Binary pulsar experiments 
constrain $\alpha_2\gtrsim-26$~\cite{Damour:1996ke}. Using the maximum 
allowed value we get 
$\mu/M\gtrsim0.1 (r_0/M)^2$, for $l=0$ and $n\gg1$.
Note that it is the combination $\alpha_2 T^E$ that regulates 
the instability. If some exotic form of matter such that $T^E>0$ surrounds the BH, then the instability occurs for 
\emph{positive} values of $\alpha_2$, which are not constrained by 
observations.

For consistency, the result above requires $\mu\ll M$ in order for Schwarzschild to be a background solution even in presence of matter. It might seem hard to be within the range of validity of this approximation and still satisfy the inequality~\eqref{model1}. 
However, the instability is quite generic and occurs also for consistent background solutions, as we now show. Consider a spherically symmetric BH -- described by the Schwarzschild geometry -- endowed with a spherical thin-shell at some radius $R$,
\be
ds^2=-h(r)dt^2+f(r)^{-1}dr^2+r^2d\Omega^2\,,\label{metric_ansatz}
\ee
where $h(r)=f(r)=1-2M/r$ for $r>R$ and $h(r)=f(r)=1-2M_{\rm int}/r$ for $r<R$. This is an exact solution of the field equations. Once the surface energy density $\sigma$ and pressure $P$ are specified, Israel's junction conditions~\cite{Israel:1966rt} provide the internal mass $M_{\rm int}$ and the shell location $R$ in terms of $\sigma$, $P$ and $M$. In this case, the sufficient condition~\eqref{sufficient} becomes:
\begin{equation}
2\alpha_2(2P-\sigma)>\frac{2l (l+1)+1}{4 M_{\rm int}}+\frac{M-M_{\rm int}}{R^2}>0\,.
\end{equation}
Therefore, if $\sigma>2P$ scalarization may occur if $\alpha_2$ is sufficiently negative whereas, if the strong energy condition is violated and $\sigma<2P$, the instability occurs for large enough values of $\alpha_2>0$.

The above models consider spherically symmetric matter distributions, but the effect is very generic. 
By expanding a generic matter distribution as $\mu_s^2(r,\theta,\phi)=\sum_{lm} \mu_{s\,lm}^2(r)Y_{lm}(\theta,\phi)$, it is easy to show that the monopole $\Psi_{00}$ decouples form the higher harmonics and satisfies Eq.~\eqref{KGsphe} with $\mu_s^2\to\mu_{s\,00}^2/\sqrt{4\pi}$.
We conclude that scalarization must occur \emph{at least} at the level of the $l=0$ mode. 
Finally, we found that spontaneous scalarization is active also when the BH rotates~\cite{Cardoso:2013opa}. 

\noindent{\bf{\em  The final state of spontaneous scalarization.}}
%
\begin{figure}
\begin{center}
\begin{tabular}{c}
\epsfig{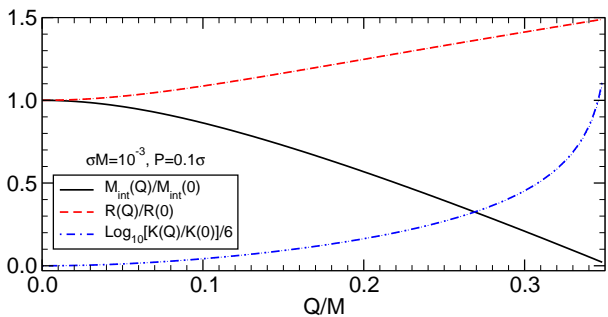}
\end{tabular}
\end{center}
\caption{\label{fig:shell}
The internal mass, $M_{\rm int}\equiv M-C/2$, shell radius $R$ and Kretschmann scalar $K=R_{abcd}R^{abcd}$ at the horizon for a hairy BH as function of the scalar charge $Q$, normalized by their value in GR, $Q=0$. 
}
\end{figure}
To understand the development of the instability and the approach to the final state, a nonlinear
time evolution is mandatory. However, interesting information on the final state
can be obtained by looking at stationary solutions of the field equations with the same symmetries.
Let us work out the spontaneously scalarized final state for thin-shell of matter surrounding a BH in spherical symmetry.
Spacetime is described again by \eqref{metric_ansatz}. For a zero-thickness shell, the matter content is zero everywhere and the Klein-Gordon equation can be integrated to yield $\Phi'=Q/(r^2\sqrt{fh})$.
The scalar charge $Q$ can be determined as a function of the matter density $\sigma$ and pressure $P$ on the shell.
The solution for $\Phi'$ implies that if there is an horizon ($f=0$) inside the shell, regularity of the scalar field imposes $\Phi={\rm const}$ inside the shell and a Schwarzschild interior. Equation~(\ref{einein}) reduces to
\beq
4\pi Q^2+r^2 h\left(f+r f'-1\right)&=&0\,,\label{eqE1}\\
4\pi Q^2+r^2 h\left(1-f\right)-r^3 f h'&=&0\,.\label{eqE2}
\eeq
For a shell made of a layer of perfect fluid, the surface stress-energy tensor reads $S_{ab}^E=\sigma u_au_b+P(\gamma_{ab}+u_au_b)$,
where $\gamma_{ab}$ denotes the induced metric and $u_a$ is the on-shell
four-velocity.
The Israel-Darmois conditions allow one to express the jump in the extrinsic curvature as function of the shell composition~\cite{Israel:1966rt}. 
The strategy is to integrate Eqs.~\eqref{eqE1}--\eqref{eqE2} from infinity, with appropriate boundary conditions, towards the shell; then use the matching conditions to get across the shell and match onto a Schwarzschild interior.
A nonlinear solution thus constructed is shown in Fig.~\ref{fig:shell}.

A perturbative analysis for small $Q$ is perhaps more illuminating. By defining $h\equiv1-2M/r+H$ and $f\equiv 1-2M/r+F$,
in the interior $\Phi={\rm const}$ and $H=F=C/r$, whereas in the exterior we have
\beq
\Phi'&=&\frac{Q}{r(r-2M)}\,,\quad F=\frac{2\pi Q^2}{Mr} \log{\left(\frac{r}{r-2M}\right)}\,,\\
H&=&-\frac{2\pi Q^2}{M^2 r}\left[2M+ (r-M) \log\left(\frac{r-2 M}{r}\right)\right]\,.
\eeq
We imposed asymptotic flatness and $M$ is the total mass. The latter differs from the internal mass of the Schwarzschild metric, whose horizon is located at $r_h\equiv 2 M_{\rm int}=2M-C$. At large distances $\Phi\sim Q/r$. In the Jordan frame, this corresponds to a shell with an effective scalar charge $\propto Q$~\cite{Cardoso:2013opa}.
The scalar charge $Q$ is a function of $\sigma, P$ and it is determined by the Klein-Gordon equation,
%
$Q=\alpha_1(\sigma-2P)\,$,
%
where $\alpha_1$ is to be evaluated at the shell's location.
Therefore, for a given coupling $A(\Phi)$, the charge $Q$ is uniquely determined by the thermodynamical properties of the shell. Finally, for a given $Q$, the junction conditions
can be solved to get $C$ and $R$ in terms of $\sigma$ and $P$.

Once the matter content is specified, the equations above determine unambiguously 
the scalar field and the metric. In 
Fig.~\ref{fig:shell}, we show the nonlinear solution for internal mass, $M_{\rm int}\equiv 
M-C/2$, the shell radius $R$ and the Kretschmann scalar 
$K=R_{abcd}R^{abcd}$ at the BH radius as functions of the 
scalar charge $Q$. The difference to the perturbative solution
is not noticeable on the plot's scale. The perturbative solution is valid to ${\cal O}(Q^2)$ but 
the agreement is perfect also for moderately large values of $Q$, where the structure of the hairy BH can be very different from its GR counterpart.

We have thus constructed nonlinear, hairy solutions of S-T theories with a BH at the center.  
Because this is the only spherically symmetric solution to Einstein equations with a spherical matter shell, it {\it must} be the end state of the instability of a Schwarzschild BH with the same ADM mass. It would be interesting to follow the nonlinear time-dependency of the instability and the dynamical approach to this kind of nonlinear solutions.

\noindent{\bf{\em Superradiant amplification and instability.}}
When $\mu_s^2(r)>0$, spontaneous scalarization does not occur. However, a positive effective mass
squared raises the interesting prospect that a ``spontaneous superradiant instability'' is present for rotating BHs, similarly to the case of massive Klein-Gordon fields~\cite{Damour:1976kh,Detweiler:1980uk,Cardoso:2004nk,Cardoso:2005vk,Dolan:2007mj}. These two instabilities
are different in nature and, in principle, lead to two very distinct end states. The superradiant instability requires an ergosphere and is expected to terminate in a GR solution with constant scalar field and lower BH spin, while spontaneous scalarization gives rise to a nontrivial scalar profile even around static BHs.
\begin{table}[hbt]
\centering \caption{The gain coefficient for scattering of scalar waves in a matter profile ${\cal G}=\beta\Theta[r - r_0](r-r_0)/r^3$.} 
\vskip 12pt
\begin{tabular}{@{}cccccc@{}}
\hline \hline
&\multicolumn{5}{c}{${\rm Flux}_{\rm out}/{\rm Flux}_{\rm in}-1(\%)$}\\ \hline
$r_0$              &$\beta=500$  &$\beta=1000$   &$\beta=2000$            &$\beta=4000$        & $\beta=8000$\\
\hline \hline
$5.7$              &0.441            &0.604          &1.332                  &9.216                  &5.985$\times 10^{4}$\\
$6.0$              &0.415        &0.539          &1.059                   &5.589                 &513.2  \\
$10$               &0.369        &0.372          &0.380                   &0.399                 &0.825  \\
\hline \hline
\end{tabular}
\label{tab:amp}
\end{table}

We now show that spontaneous superradiant instabilities are also a generic effect of S-T theories, and 
perhaps more surprisingly that superradiant amplification of waves can increase by several orders of magnitude in these
theories. 
For simplicity, we look for separable solutions of the Klein-Gordon equation with $\varphi=\Psi(r)S(\theta)e^{-i\omega t+im\phi}$
which forces the matter profile to have the general form~\cite{Cardoso:2013opa}
\be
\mu_s^2(r,\theta)=\mu_0^2+2\frac{{\cal F}(\theta)+{\cal G}(r)}{a^2+2r^2+a^2\cos2\theta}\,.\label{separable}
\ee
The term $\mu_0$ plays the role of the canonical mass term of a massive scalar, whereas $\mu_s$ is the effective mass.
We get the following coupled system of equations:
\beq
&&\frac{(\sin\theta S')'}{\sin\theta}+\left[
a^2\left(\omega^2-\mu_0^2\right)\cos^2\theta-\frac{m^2}{\sin^2\theta}-
{\cal F}+\lambda \right]S=0,\nn\\
&&\Delta 
\frac{d}{dr}\left(\Delta\frac{d\Psi}{dr}\right)+\left[
K^2-\Delta\left({\cal G}+r^2\mu_0^2+B\right)\right]\Psi=0\,,\nn
\eeq
where $\Delta=r^2+a^2-2Mr,\,K=(r^2+a^2)\omega-am,\,B=\lambda+a^2\omega^2-2am\omega$ and $\lambda$ is a separation constant, found by imposing regularity on the angular wavefunction $S(\theta)$.

For concreteness, let us focus on a specific case of Eq.~\eqref{separable}:
 $\mu^2_{0}=0\,, {\cal F}=0\,,{\cal G}=\beta\Theta[r - r_0](r-r_0)/r^3$, and start by analyzing superradiant scattering of monochromatic waves. We show in Table \ref{tab:amp} the gain in flux as a result of inputing
a flux ${\rm Flux}_{\rm in}$ at infinity, for selected values of $\beta$ and $r_0$. Note that $\beta\propto\alpha_2$ in Eq.~\eqref{effectivemass} and large positive values of $\alpha_2$ are not constrained by observations.
For small $\beta$ one recovers the standard results, with a maximum amplification of~$~0.4\%$~\cite{Press:1972zz}.
However, for certain values of $r_0,\beta$, the amplification factor can increase by six orders of magnitude or more,
making it a potentially observable effect.

\begin{center}
\begin{figure}[h]
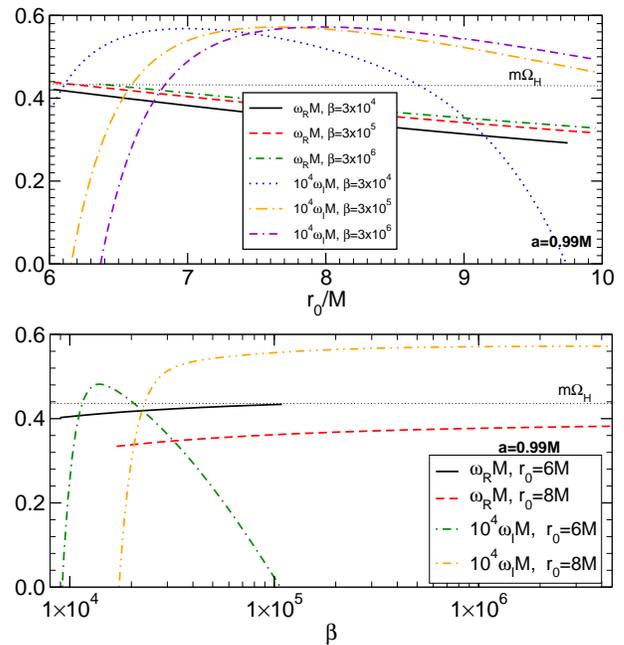

\begin{center}
\begin{tabular}{c}
\epsfig{file=modes_model_G_bis.eps,width=8cm,angle=0,clip=true}\\
\epsfig{file=modes_model_G_beta_bis.eps,width=8cm,angle=0,clip=true}
\end{tabular}
\end{center}
\caption{\label{fig:modelG}
Superradiant instability details for a matter profile characterized by ${\cal G}=\Theta[r - r_0]\beta(r-r_0)/r^3$. For large
$\beta$ the system behaves as a BH enclosed in a cavity with radius $r_0$. Curves are truncated when the modes become stable.
}
\end{figure}
\end{center}

We have also studied the full eigenvalue system to search for instabilities, which correspond to $\omega_I>0$.
Results are summarized in Fig.~\ref{fig:modelG}. 
The most important aspect to retain from our analysis is that the instability is akin to the
original BH bomb, in which a rotating BH is surrounded by a perfectly reflecting mirror at $r_0$ \cite{Press:1972zz,Cardoso:2004nk,Dolan:2012yt}:
for small $r_0$ there is no instability, as the natural frequencies of this system scale like $1/r_0$ and are outside the superradiant
regime $\omega\leq \Omega_H$, with $\Omega_H$ the BH angular velocity. It is clear from Fig.~\ref{fig:modelG} that this is a superradiant phenomenon, as the instability is quenched
as soon as one reaches the critical superradiance threshold. At fixed large $r_0/M$, and for {\it any} sufficiently large $\beta$, the instability timescale $\omega_I^{-1}$ is roughly constant. Again, in line with the simpler BH bomb system, a critical $\beta$ corresponds to a critical barrier height which is able to reflect radiation back. After this point increasing $\beta$ further is equivalent to a further increase of the height of the barrier and has no effect on the instability.

Spontaneous superradiant instability seems to be a generic feature~\cite{Cardoso:2013opa}. We have investigated also matter profiles $\mu^2_{0}\neq 0\,,{\cal F}={\cal G}=0$, 
and $\mu^2_{0}={\cal F}=0\,,{\cal G}=\mu^2r^2$ and they are equivalent or very similar to the well-known massive scalar field instability~\cite{Damour:1976kh,Detweiler:1980uk,Cardoso:2004nk,Cardoso:2005vk,Dolan:2007mj}. However, the ansatz~\eqref{separable} is not general enough and further investigation is necessary in order to understand realistic configurations 
such as accretion disks. In that case, methods such as those used in Ref.~\cite{Pani:2012vp,Pani:2012bp,Witek:2012tr,Dolan:2012yt} would be required.

\noindent{\bf{\em Conclusions.}}
BHs surrounded by matter in S-T theories are generically subjected to two instabilities. 
Spontaneous scalarization can occur when the effective mass squared is negative and, it is a very generic effect that affects GR solutions when
there is sufficient matter on the outskirts of the event horizon. The spacetime then spontaneously
develops  nontrivial scalar hair supported on the exterior matter profile. When the effective mass squared is positive superradiant instability and/or impressive amplification factors can occur.
The effectiveness of the instability depends on the matter profile, the spin of the BH and on the specific S-T theory considered.

Our results raise a number of questions, two of which are of particular interest and strongly motivate further research:
the dynamical development and final state of these instabilities; and their relevance when it comes to astrophysical BHs and potential observational imprints.

\noindent{\bf \em Acknowledgments.}
V.C. acknowledges partial financial
support provided under the European Union's FP7 ERC Starting Grant ``The dynamics of black holes:
testing the limits of Einstein's theory'' grant agreement no.
DyBHo--256667, the NRHEP 295189 FP7-PEOPLE-2011-IRSES Grant, and FCT-Portugal through projects CERN/FP/116341/2010 and CERN/FP/123593/2011.
P.P. acknowledges financial support provided by the European Community 
through the Intra-European Marie Curie contract aStronGR-2011-298297. T.P.S acknowledges financial
support from the European Research Council under the European Union's Seventh
Framework Programme (FP7/2007-2013) / ERC grant agreement n.~306425 ``Challenging
General Relativity" and from the Marie Curie Career Integration Grant LIMITSOFGR-2011-TPS.
Research at Perimeter Institute is supported by the Government of Canada 
through Industry Canada and by the Province of Ontario through the Ministry
of Economic Development and Innovation.


\end{document}